\documentclass[12pt,aps,prb,preprint]{revtex4}   
\usepackage{amsmath}    
\usepackage{amsfonts}  
\usepackage{amssymb}
\usepackage{graphicx}   

\newcommand{\ron}{{\cal R}_{\mathrm{ON}}}
\newcommand{\roff}{{\cal R}_{\mathrm{OFF}}}
\newcommand{\ro}{{\cal R}_0}
\newcommand{\rF}{{\cal R}_F}

\begin{document}

\title{The elusive memristor: properties of basic electrical circuits}

\author{Yogesh N. Joglekar}
\affiliation{Department of Physics, 
Indiana University Purdue University Indianapolis, 
Indianapolis, Indiana 46202}
\email{yojoglek@iupui.edu}   
\author{Stephen J. Wolf}
\affiliation{Department of Physics, 
Indiana University Purdue University Indianapolis, 
Indianapolis, Indiana 46202}
\date{\today}

\begin{abstract}
We present a tutorial on the properties of the new ideal circuit element, a 
memristor. By definition, a memristor $M$ relates the charge $q$ and the 
magnetic flux $\phi$ in a circuit, and complements a resistor $R$, a 
capacitor $C$, and an inductor $L$ as an ingredient of ideal electrical 
circuits. The properties of these three elements and their circuits are a 
part of the standard curricula. The existence of the memristor as the fourth 
ideal circuit element was predicted in 1971 based on symmetry arguments, but 
was clearly experimentally demonstrated just this year. We present the 
properties of a single memristor, memristors in series and parallel, as well 
as ideal memristor-capacitor (MC), memristor-inductor (ML), and 
memristor-capacitor-inductor (MCL) circuits. We find that the memristor 
has hysteretic current-voltage characteristics. We show that the ideal MC (ML) 
circuit undergoes non-exponential charge (current) decay with two time-scales, 
and that by switching the polarity of the capacitor, an ideal MCL 
circuit can be tuned from overdamped to underdamped. We present simple 
models which show that these unusual properties are closely 
related to the memristor's internal dynamics. This tutorial complements 
the pedagogy of ideal circuit elements ($R,C$, and $L$) and the properties of 
their circuits. 

\end{abstract}
\maketitle


\section{Introduction}
\label{sec:intro}

The properties of basic electrical circuits, constructed from three ideal 
elements, a resistor, a capacitor, an inductor, and an ideal 
voltage source $v(t)$, are a standard staple of physics and 
engineering courses. These circuits show a wide variety of phenomena such as 
the exponential charging and discharging of a resistor-capacitor (RC) 
circuit with time constant $\tau_{RC}=RC$, the exponential rise and 
decay of the current in a resistor-inductor (RL) circuit with time constant 
$\tau_{RL}=L/R$, the non-dissipative oscillations in an inductor-capacitor 
(LC) circuit with frequency $\omega_{LC}=1/\sqrt{LC}$, as well as resonant 
oscillations in a resistor-capacitor-inductor (RCL) circuit induced by an 
alternating-current (AC) voltage source with frequency 
$\omega\sim\omega_{LC}$.~\cite{ugbooks} The behavior of these ideal circuits 
is determined by Kirchoff's current law that follows from the continuity 
equation, and Kirchoff's voltage law. We remind the Reader that Kirchoff's 
voltage law follows from Maxwell's second equation only when the 
time-dependence of the magnetic field created by the current in the circuit 
is ignored, 
$\oint {\bf E}\cdot{\bf dl}=0$ where the line integral of the electric 
field ${\bf E}$ is taken over any closed loop in the circuit.~\cite{feynman2} 
The study of elementary circuits with ideal elements provides us 
with a recipe to understand real-world circuits where every capacitor has 
a finite resistance, every battery has an internal resistance, and every 
resistor has an inductive component; we assume that the real-world circuits 
can be modeled using only the three ideal elements and an ideal voltage 
source. 

\begin{figure}[h]
\begin{center}
\includegraphics{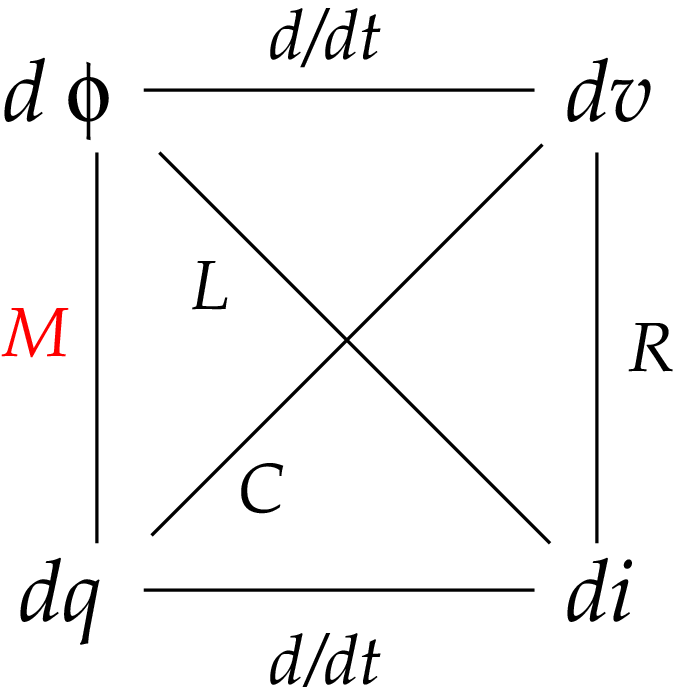}
\caption{\label{fig:square}
Relations between four variables of basic electrical circuit theory: the 
charge $q$, current $i$, voltage $v$ and the magnetic flux $\phi$. Three 
well-known ideal circuit elements $R,C$ and $L$ are associated with pairs 
($dv,di$), ($dq,dv$) and ($d\phi,di$) respectively. The top (bottom) 
horizontal pair is related by Lenz's law (definition). This leaves the pair 
$(d\phi,dq$) unrelated. Leon Chua postulated that, due to symmetry, a 
fourth ideal element (memristor) that relates this pair, $d\phi=Mdq$, must 
exist.}
\end{center}
\end{figure}

An ideal capacitor is defined by the single-valued relationship between the 
charge $q(t)$ and the voltage $v(t)$ via $dq=Cdv$. Similarly, an ideal 
resistor is defined by a single-valued relationship between the current 
$i(t)$ and the voltage $v(t)$ via $dv=Rdi$, and an ideal inductor is defined 
by a single-valued relationship between the magnetic flux $\phi(t)$ and the 
current $i(t)$ via $d\phi=Ldi$. These three definitions provide three 
relations between the four fundamental constituents of the circuit theory, 
namely the charge $q$, current $i$, voltage $v$, and 
magnetic flux $\phi$ (See Figure~\ref{fig:square}). The definition of 
current, $i=dq/dt$, and the Lenz's law, $v=+d\phi/dt$, give two more 
relations between the four constituents. (We define the flux such that the 
sign in the Lenz's law is positive). These five relations, shown in 
Fig.~\ref{fig:square}, raise a natural question: Why is an element 
relating the charge $q(t)$ and magnetic flux $\phi(t)$ missing? Based on 
this symmetry argument, in 1971 Leon Chua postulated that a new ideal 
element defined by the single-valued relationship $d\phi=Mdq$ must exist. 
He called this element memristor $M$, a short for 
memory-resistor.~\cite{chua71} This ground-breaking hypothesis meant that the 
trio of ideal circuit elements (R,C,L) were not sufficient to model a basic 
real-world circuit (that may have a memristive component as well). 
In 1976, Leon Chua and Sung Kang extended the analysis further to memristive 
systems.~\cite{chua76,chua80} These seminal articles studied the properties of 
a memristor, the fourth ideal circuit element, and showed that diverse 
systems such as thermistors, Josephson junctions, and ionic transport in 
neurons, described by the Hodgkins-Huxley model, are special cases of 
memristive systems.~\cite{chua71,chua76,chua80} 

Despite the simplicity and the soundness of the symmetry argument that 
predicts the existence of the fourth ideal element, experimental realization 
of a quasi-ideal memristor - defined by the single-valued relationship 
$d\phi=Mdq$ - remained elusive.~\cite{oldworkprogram,thakoor,ero1,ero2} 
Early this year, 
Strukov and co-workers~\cite{strukov} created, using a nano-scale thin-film 
device, the first realization of a memristor. They presented an elegant 
physical model in which the memristor is equivalent to a time-dependent 
resistor whose value at time $t$ is linearly proportional to the amount of 
charge $q$ that has passed through it before. This equivalence follows 
from the memristor's definition and Lenz's law, 
$d\phi=Mdq\Leftrightarrow v=M(q)i$. It also implies that the memristor value - 
memristance - is measured in the same units as the resistance. 

In this tutorial, we present the properties of basic electrical circuits 
with a memristor. For the most part, this theoretical investigation uses 
Kirchoff's law and Ohm's law. In the next section, we discuss the memristor 
model presented in Ref.~\onlinecite{strukov} and analytically derive its 
{\it i-v} characteristics. Section~\ref{sec:mc} contains theoretical 
results for ideal MC and ML circuits. We use the linear drift model, presented 
in Ref.~\onlinecite{strukov}, to describe the dependence of the effective 
resistance of the memristor (memristance) on the charge that has passed 
through it. This simplification allows us to obtain analytical closed-form 
results. We show 
the charge (current) decay ``time-constant'' in an ideal MC (ML) circuit 
depends on the polarity of the memristor. Sec.~\ref{sec:nldrift} is intended 
for advanced students. In this section, we present models that 
characterize the dependence of the memristance on the dopant drift inside 
the memristor. We show that the memristive behavior is amplified when we use 
models that are more realistic than the one used in preceding sections. 
In Sec.~\ref{sec:mcl} we discuss an ideal MCL circuit. We show that 
depending on the polarity of the memristor, the MCL circuit can be 
overdamped or underdamped, and thus allows far more tunability than an 
ideal RCL circuit.  Sec.~\ref{sec:disc} concludes the tutorial with a 
brief discussion. 


\section{A Single Memristor}
\label{sec:sm}

\begin{figure}[h]
\begin{center}
\scalebox{0.8}{\includegraphics{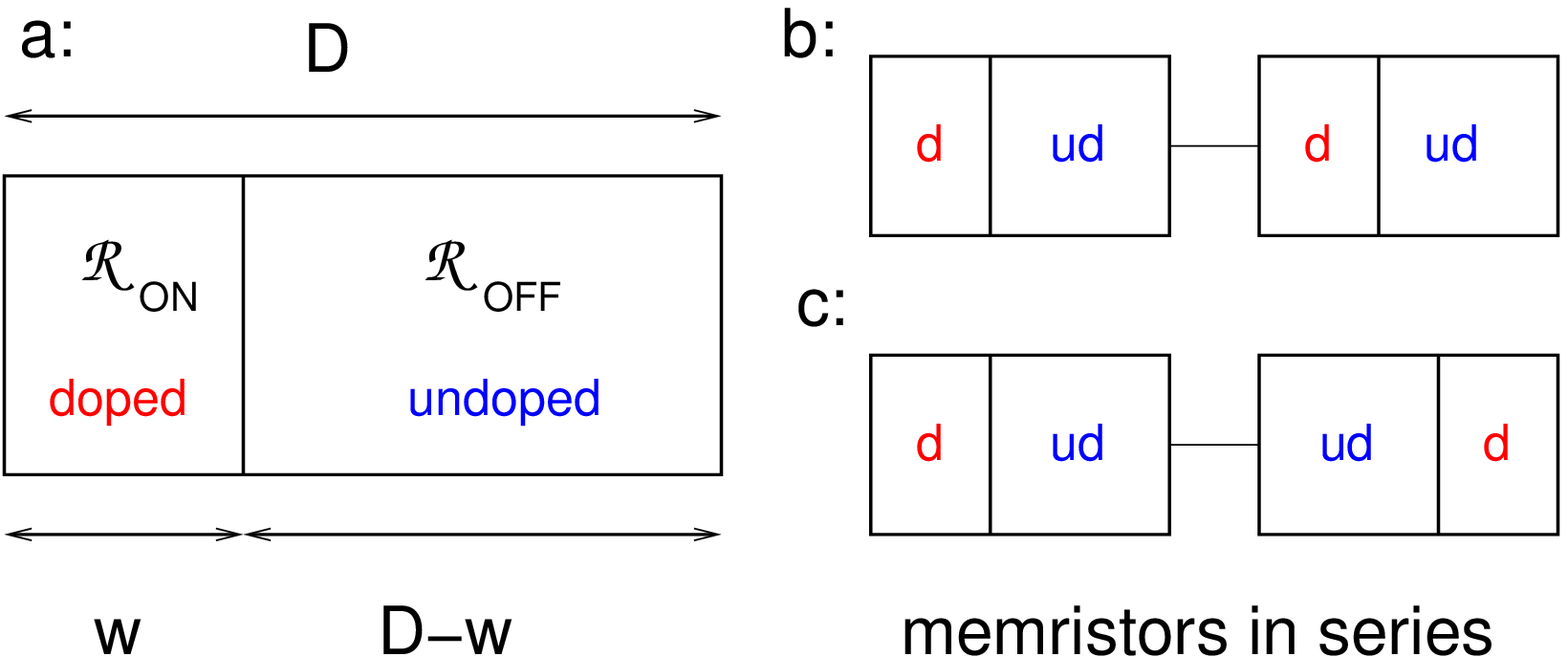}}
\caption{\label{fig:schematic}
a: Schematic of a memristor of length $D$ as two resistors in series. The 
doped region (TiO$_{2-x}$ in Ref.~\onlinecite{strukov}) has resistance 
$\ron w/D$ and the undoped region (TiO$_2$ in Ref.~\onlinecite{strukov}) 
has resistance $\roff(1-w/D)$. The size of the doped region, with its 
charge +2 ionic dopants, changes in response to the applied voltage and 
thus alters the effective resistance of the memristor. 
b: Two memristors with the same polarity in 
series. {\sf d} and {\sf ud} represent the doped and undoped regions 
respectively. In this case, the memristive effect is retained because doped 
regions in both memristors simultaneously shrink or expand. 
c: Two memristors with opposite polarities in series. The net memristive 
effect is suppressed.}
\end{center}
\end{figure}

We start this section with the elegant model of a memristor presented in 
Ref.~\onlinecite{strukov}. It consisted of a thin film (5 nm thick) with 
one layer of insulating TiO$_2$ and oxygen-poor TiO$_{2-x}$ each, 
sandwiched between platinum contacts. The oxygen vacancies in the second 
layer behave as charge +2 mobile dopants. These dopants create a doped 
TiO$_2$ region, whose resistance is significantly lower than the 
resistance of the undoped region. The boundary between the doped and 
undoped regions, and therefore the effective resistance of the thin film, 
depends on the position of these dopants. It, in turn, is determined by 
their mobility $\mu_D$ ($\sim 10^{-10}$ cm$^2$/V.s)~\cite{strukov} and the 
electric field across the doped region.~\cite{ugbooks1} 
Figure~\ref{fig:schematic} shows a schematic of a memristor of size $D$ 
($D\sim$ 10 nm) modeled as two resistors in series, the doped region with 
size $w$ and the undoped region with size $(D-w)$. The effective resistance 
of such a device is 
\begin{equation}
\label{eq:memdef}
M(w)=\frac{w}{D}\ron+\left(1-\frac{w}{D}\right)\roff
\end{equation}
where $\ron$ ($\sim$1k$\Omega$)~\cite{strukov} is the resistance of the 
memristor if it is completely doped, and $\roff$ is its resistance if it is 
undoped. Although Eq.(\ref{eq:memdef}) is valid for arbitrary values of 
$\ron$ and $\roff$, experimentally, the resistance of the doped TiO$_2$ film 
is significantly smaller than the undoped film, $\roff/\ron\sim 10^2\gg 1$ and 
therefore $\Delta{\cal R}=(\roff-\ron)\approx\roff$. In the presence 
of a voltage $v(t)$ the current in the memristor is determined by Kirchoff's 
voltage law $v(t)=M(w)i(t)$. The memristive behavior of this system is 
reflected in 
the time-dependence of size of the doped region $w(t)$. In the simplest 
model - the linear-drift model - the boundary between the doped and the 
undoped regions drifts at a constant speed $v_D$ given by 
\begin{equation}
\label{eq:driftv}
\frac{dw}{dt}=v_D=\eta\frac{\mu_D\ron}{D}i(t)
\end{equation}
where we have used the fact that a current $i(t)$ corresponds to a uniform 
electric field $\ron i(t)/D$ across the doped region. Since the (oxygen 
vacancy) dopant drift can either expand or contract the doped region, we 
characterize the ``polarity'' of a memristor by $\eta=\pm1$, where $\eta=+1$ 
corresponds to the expansion of the doped region. We note that ``switching 
the memristor polarity'' means reversing the battery terminals, or the $\pm$ 
plates of a capacitor (in an MC circuit) or reversing the direction of the 
initial current (in an ML circuit). Eqns.(\ref{eq:memdef})-(\ref{eq:driftv}) 
are used to determine the {\it i-v} characteristics of a memristor. 
Integrating Eq.(\ref{eq:driftv}) gives
\begin{equation}
\label{eq:w}
w(t)=w_0+\eta\frac{\mu_D\ron}{D}q(t)=w_0+\eta\frac{Dq(t)}{Q_0}
\end{equation}
where $w_0$ is the initial size of the doped region. Thus, the width of the 
doped region $w(t)$ changes linearly with the amount of charge that has 
passed through it.~\cite{caveat1} $Q_0=D^2/\mu_D\ron$ is the charge that is 
required to pass through the memristor for the dopant boundary to move 
through distance $D$ (typical parameters~\cite{strukov} imply 
$Q_0\sim 10^{-2}$ C). It provides the natural scale for charge in a 
memristive circuit. Substituting this result in Eq.(\ref{eq:memdef}) gives  
\begin{equation}
\label{eq:memformula}
M(q)=\ro-\eta\frac{\Delta{\cal R}q}{Q_0},  
\end{equation}
where $\ro=\ron(w_0/D)+\roff(1-w_0/D)$ is the effective resistance 
(memristance) at time $t=0$. Eq.(\ref{eq:memformula}) shows explicitly that 
the memristance $M(q)$ depends purely on the charge $q$ that has passed 
through it. Combined with $v(t)=M(q)i(t)$, Eq.(\ref{eq:memformula}) implies 
that the model presented here is an ideal memristor. (We recall that 
$v=M(q)i$ is equivalent to $d\phi=Mdq$). The prefactor of the $q$-dependent 
term is proportional to $1/D^2$ and becomes increasingly important when $D$ 
is small. In addition, for a given $D$, the memristive effects become 
important only when $\Delta{\cal R}\gg\ro$. Now that we have discussed the 
memristor model from Ref.~\onlinecite{strukov}, in the following paragraphs 
we obtain analytical results for its {\it i-v} characteristics. 

\begin{figure}[t]
\begin{center}
\vspace{-12cm}
\includegraphics{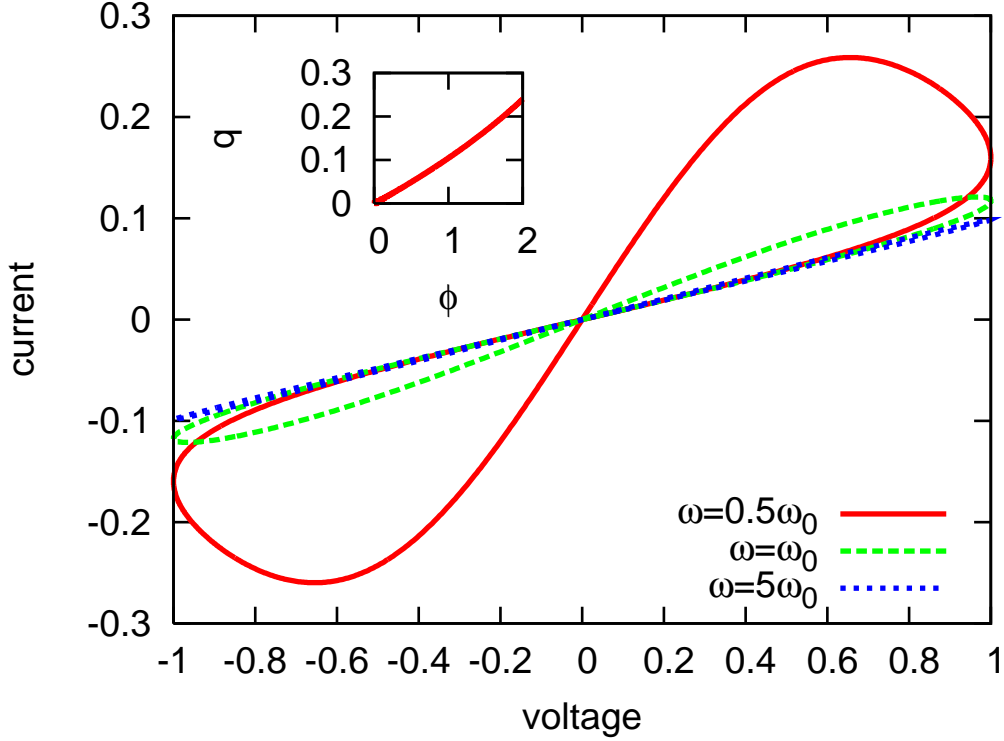}
\caption{\label{fig:hyste}
Theoretical {\it i-v} characteristics of a memristor with applied voltage 
$v(t)=v_0\sin(\omega t)$ for $\omega=0.5\omega_0$ (red solid), 
$\omega=\omega_0$ (green dashed), and $\omega=5\omega_0$ (blue dotted). The 
memristor parameters are $w_0/D=0.5$ and $\roff/\ron=20$. The unit 
of resistance is $\ron$, the unit of voltage is $v_0$, and the unit of 
current is $I_0=Q_0/t_0$. We see that the hysteresis is pronounced for 
$\omega\le\omega_0$ and suppressed when $\omega\gg\omega_0$. The inset is 
a typical {\it q-$\phi$} graph showing that the charge $q$ is an invertible 
function of the flux $\phi$. The unit of flux $\phi_0=v_0t_0=D^2/\mu_D$ is 
determined by the memristor properties alone (typical 
parameters~\cite{strukov} imply $\phi_0=10^{-2}$ Wb).} 
\end{center}
\end{figure}

For an ideal circuit with a single memristor and a voltage supply, Kirchoff's 
voltage law implies
\begin{equation}
\label{eq:memlinear}
\left(\ro-\eta\frac{\Delta{\cal R}q(t)}{Q_0}\right)\frac{dq}{dt}=
\frac{d}{dt}\left(\ro q-\eta\frac{\Delta{\cal R}q^2}{2Q_0}\right)=v(t).  
\end{equation}
The solution of this equation, subject to the boundary condition $q(0)=0$ is 
\begin{eqnarray}
\label{eq:charge}
q(t)& = & \frac{Q_0\ro}{\Delta{\cal R}}
\left[1-\sqrt{1-\eta\frac{2\Delta{\cal R}}{Q_0\ro^2}\phi(t)}\right], \\
\label{eq:current}
i(t)& = & \frac{v(t)}{\ro}\frac{1}{\sqrt{1-2\eta\Delta{\cal R}\phi(t)/
Q_0\ro^2}}=\frac{v(t)}{M(q(t))},
\end{eqnarray}
where $\phi(t)=\int_0^{t}d\tau v(\tau)$ is the magnetic flux associated with 
the voltage $v(t)$. Eqs.(\ref{eq:charge})-(\ref{eq:current}) provide 
analytical results for {\it i-v} characteristics of an ideal memristor 
circuit. Eq.(\ref{eq:charge}) shows that the charge is an invertible function 
of the magnetic 
flux~\cite{chua71,chua76} consistent with the defining equation 
$d\phi=M(q)dq$. Eq.(\ref{eq:current}) shows that a memristor does not 
introduce a phase-shift between the current and the voltage, $i=0$ if and 
only if $v=0$. Therefore, unlike an ideal capacitor or an inductor, is a 
purely dissipative element.~\cite{chua71} For an AC voltage 
$v(t)=v_0\sin(\omega t)$, the magnetic flux is 
$\phi(t)=v_0[1-\cos(\omega t)]/\omega$. Note that although 
$v(\pi/\omega-t)=v(t)$, $\phi(\pi/\omega-t)\neq \phi(t)$. Therefore, it 
follows from Eq.(\ref{eq:current}) that the current $i(v)$ will be a 
multi-valued function of the voltage $v$. It also follows that since 
$\phi\propto 1/\omega$, the memristive behavior is dominant only at low 
frequencies $\omega\lesssim\omega_0=2\pi/t_0$. Here $t_0=D^2/\mu_Dv_0$ is 
the time that the dopants need to travel distance $D$ under a constant 
voltage $v_0$. $t_0$ and $\omega_0$ provide the natural time and frequency 
scales for a memristive circuit (typical parameters~\cite{strukov} imply 
$t_0\sim 0.1$ ms and $\omega_0\sim 50$ KHz). We emphasize that 
Eq.(\ref{eq:charge}) is based on the linear-drift model, 
Eq.(\ref{eq:driftv}), and is valid~\cite{caveat1} only when the charge 
flowing through the memristor is less than $q_{\max}(t)=Q_0(1-w_0/D)$ 
when $\eta=+1$ or $q_{\max}(t)=Q_0w_0/D$ when $\eta=-1$. It is easy to 
obtain a diversity of {\it i-v} characteristics using 
Eqns.(\ref{eq:charge}) and (\ref{eq:current}), including those presented in 
Ref.~\onlinecite{strukov} by choosing appropriate functional forms of $v(t)$. 
Figure~\ref{fig:hyste} shows the theoretical {\it i-v} curves for 
$v(t)=v_0\sin(\omega t)$ for 
$\omega=0.5\omega_0$ (red solid), $\omega=\omega_0$ (green dashed), and 
$\omega=5\omega_0$ (blue dotted). In each case, the high initial resistance 
$\ro$ leads to the small slope of the {\it i-v} curves at the beginning. For 
$\omega\leq\omega_0$ as the voltage increases, the size of the doped region 
increases and the memristance decreases. 
Therefore, the slope of the {\it i-v} curve on the return sweep is large 
creating a hysteresis loop. The size of this loop varies inversely with 
the frequency $\omega$. At high frequencies, $\omega=5\omega_0$, the size 
of the doped region barely changes before the applied voltage begins the 
return sweep. Hence the memristance remains essentially unchanged and 
the hysteretic behavior is suppressed. The inset in Fig.~\ref{fig:hyste} 
shows the theoretical {\it q}-$\phi$ curve for $\omega=0.5\omega_0$ that 
follows from Eq.(\ref{eq:charge}).  

Thus, a single memristor shows a wide variety of {\it i-v} characteristics 
based on the frequency of the applied voltage. Since the mobility of the 
(oxygen vacancy) dopants is low, memristive effects are appreciable only 
when the memristor size is nano-scale. Now, we consider an ideal circuit 
with two memristors in series (Fig.~\ref{fig:schematic}). It follows 
from Kirchoff's laws that if two memristors $M_1$ and $M_2$ have the same 
polarity, $\eta_1=\eta_2$, they add like regular resistors, 
$M(q)=({\cal R}_{01}+{\cal R}_{02})-\eta(\Delta{\cal R}_1+
\Delta{\cal R}_2)q(t)/Q_0$ whereas when they have opposite polarities, 
$\eta_1\eta_2=-1$, the $q$-dependent component is suppressed, 
$M(q)=({\cal R}_{01}+{\cal R}_{02})-\eta(\Delta{\cal R}_1-
\Delta{\cal R}_2)q(t)/Q_0$. The fact that memristors with same polarities 
add in series leads to the possibility of a superlattice of memristors 
with micron dimensions instead of the nanoscale dimensions. We emphasize 
that a single memristor cannot be scaled up without losing the memristive 
effect because the relative change in the size of the doped region 
decreases with scaling. A superlattice of nano-scale memristors, on the 
other hand, will show the same memristive effect when scaled up. We leave 
the problem of two memristors in parallel to the Reader. 

These non-trivial properties of an ideal memristor circuit raise the 
following question: What are the properties of basic circuits with a 
memristor and a capacitor or an inductor? (A memristor-resistor circuit is 
trivial.) We will explore this question in the subsequent sections. 


\section{Ideal MC and ML Circuits}
\label{sec:mc}

\begin{figure}[h]
\begin{center}
\vspace{-12cm}
\includegraphics{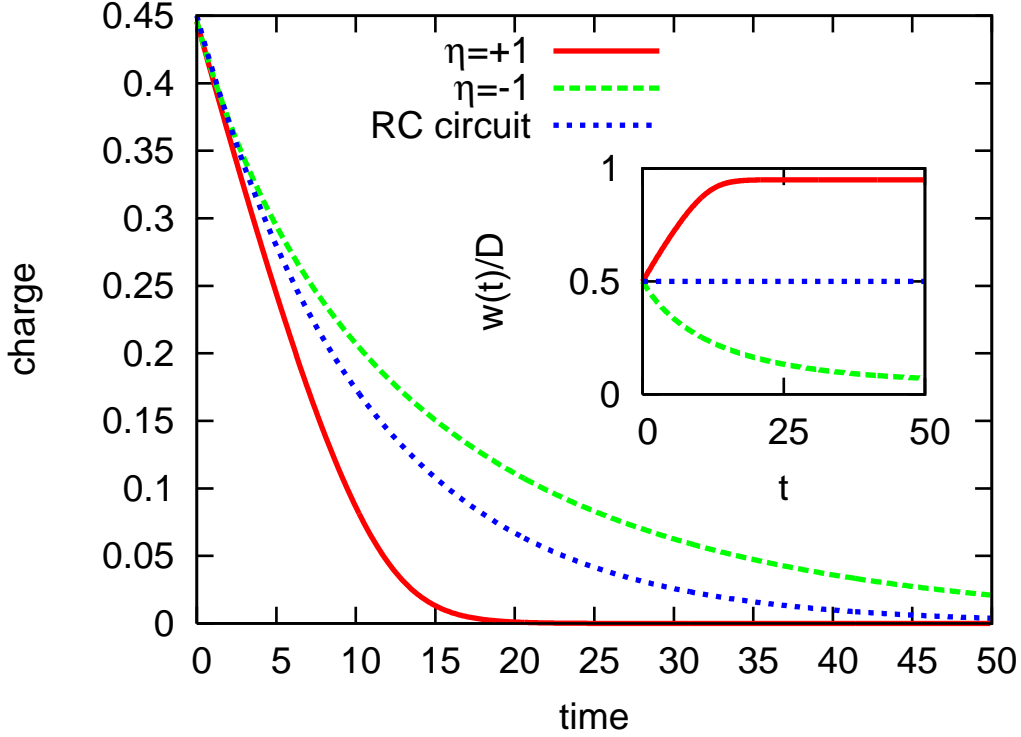}
\caption{\label{fig:mc}
Theoretical {\it q-t} characteristics of an ideal MC circuit. The memristor 
parameters are $w_0/D=0.5$ and $\roff/\ron=20$. The initial charge on the 
capacitor is $q_0/Q_0=0.45<(1-w_0/D)$ to ensure the validity of linear-drift 
model,~\cite{caveat1} and $C/C_0=1$. The unit of capacitance is 
$C_0=Q_0/v_0=t_0/\ron$. 
We see that when $\eta=+1$ (red solid), the capacitor charge in the MC 
circuit decays about {\it twice as fast as} when $\eta=-1$ (green dashed). 
The central blue dotted 
plot shows the exponential charge decay of an RC circuit with same initial 
resistance $\ro$. The inset shows the time-evolution of the boundary 
between the doped and undoped regions when $\eta=+1$ (red solid) and 
$\eta=-1$ (green dashed), and confirms that the linear-drift model 
is valid.~\cite{caveat1}}
\end{center}
\end{figure}

Let us consider an ideal MC circuit with a capacitor having an initial 
charge $q_0$ and no voltage source. The effective resistance of the 
memristor is determined by its polarity (whether the doped region increases 
or decreases), and since the charge decay time-constant of the MC circuit 
depends on its effective resistance, the capacitor discharge will depend on 
the memristor polarity. Kirchoff's voltage law applied to an ideal MC 
circuit gives  
\begin{equation}
\label{eq:mcdischarge}
M_c(q(t))\frac{dq}{dt}+\frac{q}{C}=0
\end{equation} 
where $q(t)$ is the charge on the capacitor. We emphasize that the 
$q$-dependence of the memristance here is 
$M_c(q)=\ro-\eta\Delta{\cal R}(q_0-q)/Q_0$ because if $q$ is the remaining 
charge on the capacitor, then the charge that has passed through the 
memristor is ($q_0-q$). Eq.(\ref{eq:mcdischarge}) is integrated by 
rewriting it as $dq/dt=-q/(a+bq)$ where $a=C(\ro-\eta\Delta{\cal R}q_0/Q_0)$ 
and $b=\eta C\Delta{\cal R}/Q_0$. We obtain the following implicit equation 
\begin{equation}
\label{eq:implicitq}
q(t)\exp\left[\frac{\eta\Delta{\cal R}q(t)}{\rF Q_0}\right]=q_0\exp
\left[-\frac{t}{\rF C}\right]\exp\left[\frac{\eta\Delta{\cal R}q_0}
{\rF Q_0}\right] 
\end{equation}
where $\rF=\ro-\eta\Delta{\cal R} q_0/Q_0$ is the memristance when the 
entire charge $q_0$ has passed through the memristor.~\cite{caveat1} 
A small $t$-expansion of Eq.(\ref{eq:implicitq}) shows that the initial \
current $i(0)=q_0/\ro C$ is independent of the memristor polarity $\eta$, 
and the large-$t$ expansion shows that the charge on the capacitor decays 
exponentially, $q(t\rightarrow\infty)=
q_0\exp(-t/\rF C)\exp(\eta\Delta{\cal R}q_0/\rF Q_0)$. In the intermediate 
region, the naive expectation $q(t)=q_0\exp[-t/M(w(t))C]$ is not the 
self-consistent solution of Eq.(\ref{eq:implicitq}). Therefore, although a 
memristor can be thought of as an effective resistor, its effect in an MC 
circuit is not captured by merely substituting its time-dependent value in 
place of the resistance in an ideal RC circuit. Qualitatively, since the 
memristance decreases or increases depending on its polarity, we expect 
that when $\eta=+1$ the MC circuit will discharge faster than an RC 
circuit with same resistance $\ro$. 
That RC circuit, in turn, will discharge faster than the same MC circuit 
when $\eta=-1$. Figure~\ref{fig:mc} shows the theoretical {\it q-t} curves 
obtained by (numerically) integrating Eq.(\ref{eq:mcdischarge}). These results 
indeed fulfill our expectations. We note that Eq.(\ref{eq:implicitq}), 
obtained using the linear-drift model, is valid for $q_0\leq Q_0(1-w_0/D)$ 
when $\eta=+1$ which guarantees that the final memristance $\rF\geq\ron$ 
is always positive.~\cite{caveat1} The inset in Fig.~\ref{fig:mc} shows the 
time-evolution of the size of the doped region $w(t)$ obtained using 
Eq.(\ref{eq:w}) and confirms the applicability of the linear-drift model. 
We remind the Reader that changing the polarity of the memristor can be 
accomplished by exchanging the $\pm$ plates of the fully charged capacitor. 

It is now straightforward to understand an ideal MC circuit with a 
direct-current (DC) voltage source $v_0$ and an uncharged capacitor. This 
problem is the time-reversed version of an MC circuit with the capacitor 
charge $q_0=v_0C$ and no voltage source. The only salient difference is that 
in the present case, the charge passing through the memristor is the same as 
the charge on the capacitor. Using Kirchoff's voltage law we obtain the 
following implicit result,
\begin{equation}
\label{eq:implicitq2}
q(t)=v_0C\left[1-\exp\left(-\frac{t}{\rF C}+\frac{\eta\Delta{\cal R}q(t)}
{\rF Q_0}\right)\right]
\end{equation}
where $\rF=\ro-\eta\Delta{\cal R}(v_0C)/Q_0$ is the memristance when 
$t\rightarrow\infty$. As before, Eq.(\ref{eq:implicitq2}) shows that 
when $eta=+1 (\eta=-1)$, the ideal MC circuit charges faster (slower) than 
an ideal RC circuit with the same resistance $\ro$. In particular, 
the capacitor charging time for $\eta=+1$ (the doped region widens and the 
memristance reduces with time) decreases steeply as the DC voltage 
$v_0\rightarrow Q_0(1-w_0/D)/C$, the maximum voltage at which the 
linear-drift model is applicable.~\cite{caveat1}  

\begin{figure}[h]
\begin{center}
\vspace{-12cm}
\includegraphics{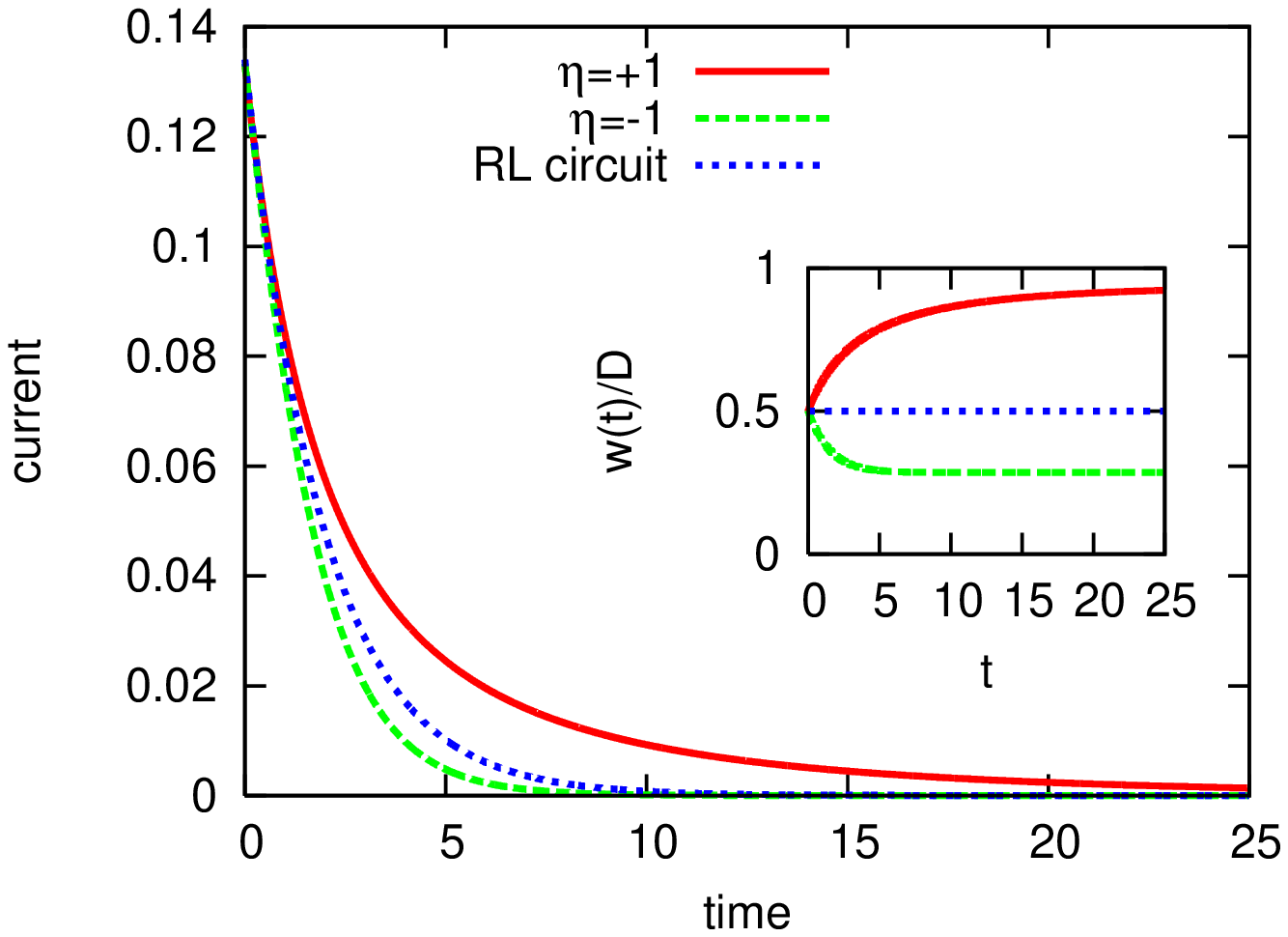}
\caption{\label{fig:ml}
Theoretical {\it i-t} characteristics of an ideal ML circuit. The memristor 
parameters are $w_0/D=0.5$ and $\roff/\ron=30$. The initial current 
in the circuit is small, $i_0/I_0=0.135$, to ensure the validity of the 
linear-drift model~\cite{caveat1} that breaks down when $i_0/I_0>0.140$, 
and $L/L_0=30$. 
The unit of inductance is $L_0=\phi_0/I_0=t_0\ron$. We see that when 
$\eta=+1$ (red solid), the current in the ML circuit decays {\it slower} 
than when $\eta=-1$ (green dashed). The central blue dotted plot shows the 
exponential current decay of an RL circuit with same initial resistance 
$\ro$. The inset shows the time-evolution of the boundary between the doped 
and undoped regions when $\eta=+1$ (red solid) and $\eta=-1$ (green dashed), 
and confirms that the linear-drift model is valid.~\cite{caveat1}}
\end{center}
\end{figure}

Now we turn our attention to an ML circuit. Ideal RC and RL circuits are 
described by the same differential equation 
($dq/dt+q/\tau_{RC}=0; di/dt+i/\tau_{RL}=0$) with same boundary conditions. 
Therefore they have identical solutions~\cite{feynman2} 
$q(t)=q_0\exp(-t/\tau_{RC})$ and $i(t)=i_0\exp(-t/\tau_{RL})$. 
As we will see below, this equivalence breaks down for MC and ML circuits. 
Let us consider an ideal ML circuit with initial current $i_0$. Kirchoff's 
voltage law implies that 
\begin{equation} 
\label{eq:ml}
Li\frac{di}{dq}+\left(\ro-\eta\frac{\Delta{\cal R}q(t)}{Q_0}\right)i(t)=0.
\end{equation}
The solution of this equation above is given by 
$i(q)=Aq^2(t)-Bq(t)+i_0=(q-q_+)(q-q_-)$ where 
$A=\eta\Delta{\cal R}/2Q_0L$, $B=\ro/L$, and 
$q_{\pm}=(Q_0\ro/\Delta{\cal R})\left[1\pm
\sqrt{1-2\eta\Delta{\cal R}Li_0/Q_0\ro^2}\right]$ are the two real roots of 
$i(q)=0$. We integrate the implicit result using partial fractions and get 
\begin{equation}
\label{eq:implicitql}
q(t)=\frac{2Q_0L i_0}{\Delta{\cal R}}\left[\frac{e^{t/\tau_{ML}}-1}
{q_{+}e^{t/\tau_{ML}}-q_{-}}\right]
\end{equation}
where $\tau_{ML}=L/\ro\sqrt{1-2\eta\Delta{\cal R}Li_0/Q_0\ro^2}$ is 
characteristic time associated with the ML circuit. The current $i(t)$ in 
the circuit is 
\begin{equation}
\label{eq:currentl}
i(t)=i_0\left(\frac{2Q_0L}{\Delta{\cal R}\tau_{ML}}\right)^2
\frac{e^{t/\tau_{ML}}}{(q_{+}e^{t/\tau_{ML}}-q_{-})^2}. 
\end{equation}
Eqs.(\ref{eq:implicitql})-(\ref{eq:currentl}) provide the set of 
analytical results for an ideal ML circuit. At small-$t$ 
Eq.(\ref{eq:currentl}) becomes $i(t)=i_0(1-t\ro/L)$, whereas the 
large-$t$ expansion shows that the current decays exponentially, 
$i(t\rightarrow\infty)=i_0(2Q_0L/
q_{+}\Delta{\cal R}\tau_{ML})^2\exp(-t/\tau_{ML})$. Since $\tau_{ML}$ depends 
on the polarity of the memristor, $\tau_{ML}(\eta=+1)>\tau_{ML}(\eta=-1)$, 
the ML circuit with $\eta=+1$ discharges slower than its RL counterpart 
whereas the same ML circuit with $\eta=-1$ discharges faster than the RL 
counterpart. Figure~\ref{fig:ml} shows the theoretical {\it i-t} curves for 
an ML circuit obtained from Eq.(\ref{eq:currentl}); these results are 
consistent with our qualitative analysis. Note that the net charge passing 
through the memristor in an ML circuit is $q(t\rightarrow\infty)=q_{-}(i_0)$. 
Therefore an upper limit on initial current $i_0$ for the validity of the 
linear-drift model~\cite{caveat1} is given by 
$q_{-}(i_0)\leq Q_0w_0/D$ ($\eta=-1$). As in the case of an ideal MC 
circuit charge, the ML circuit current decays steeply as $i_0$ approaches this 
upper limit. 

Figs.~\ref{fig:mc} and~\ref{fig:ml} suggest that ideal MC and ML circuits 
have a one-to-one correspondence analogous to the ideal RC and RL circuits. 
Therefore, it is tempting to think that solution of an ideal ML circuit 
with a DC voltage $v_0$ is straightforward. (In a corresponding RL circuit, 
the current asymptotically approaches $v_0/R$ for $t\gg\tau_{RL}=L/R$). The 
relevant differential equation obtained using Kirchoff's voltage law,
\begin{equation}
\label{eq:mlv}
L\frac{di}{dt}+\left(\ro-\eta\frac{\Delta{\cal R}q(t)}{Q_0}\right)i(t)=v_0, 
\end{equation}
shows that it is not the case. In an ML circuit, as the current $i(t)$ 
asymptotically approaches its maximum value, it can pump an arbitrarily 
large charge $q(t)=\int^{t}_0i(\tau)d\tau$ through the memristor. Hence, for 
any non-zero voltage, no matter how small, the linear-drift model breaks 
down at large times when $w(t)=w_0+\eta Dq(t)/Q_0$ exceeds $D$ ($\eta=+1$) 
or becomes negative ($\eta=-1$). This failure of the linear-drift model 
reflects the fact that when the (oxygen vacancy) dopants approach either 
end of the memristor, their drift is strongly suppressed by a non-uniform 
electric field. Thus, unlike the ideal RL circuit, 
the steady-state current in an ideal ML circuit is not solely determined by 
the resistance $\ro$ but also by the inductor. In the following section, we 
present more realistic models of the dopant drift that take into account its 
suppression near the memristor boundaries. 


\section{Models of Non-linear Dopant Drift}
\label{sec:nldrift}

The linear-drift model used in preceding sections captures the majority of 
salient features of a memristor. It makes the ideal memristor, MC, and ML 
circuits analytically tractable and leads to closed-form results such as 
Eqs.~(\ref{eq:current}), (\ref{eq:implicitq}), and (\ref{eq:currentl}). We 
leave it as an exercise for the Reader to verify that these results reduce 
to their well-known R, RC, and RL counterparts in the limit when the 
memristive effects are negligible, $\Delta{\cal R}\rightarrow 0$. The linear 
drift model suffers from one serious drawback: it does not take into 
account the boundary effects. Qualitatively, the boundary between the doped 
and undoped regions moves with speed $v_D$ in the bulk of the memristor, but 
that speed is strongly suppressed when it approaches either edge, $w\sim 0$ 
or $w\sim D$. We modify Eq.(\ref{eq:driftv}) to reflect this suppression as 
follows~\cite{strukov} 
\begin{equation}
\label{eq:nldriftv}
\frac{dw}{dt}=\eta\frac{\mu_D\ron}{D}i(t)F\left(\frac{w}{D}\right). 
\end{equation}
The window function $F(x)$ satisfies $F(0)=F(1)=0$ to ensure no drift at the 
boundaries. The function $F(x)$ is symmetric about $x=1/2$ and monotonically 
increasing over the interval $0\leq x\leq 1/2$, $0\leq F(x)\leq 1=F(x=1/2)$. 
These properties guarantee that the difference between this model and the 
linear-drift model, Eq.(\ref{eq:driftv}), vanishes in the bulk of the 
memristor as $w\rightarrow D/2$. Motivated by this physical picture, we 
consider a family of window functions parameterized by a positive integer 
$p$, $F_p(x)=1-(2x-1)^{2p}$. 
Note that $F_p(x)$ satisfies all the constraints for any $p$. The equation 
$F_{p}(x)=0$ has 2 real roots at $x=\pm 1$, and $2(p-1)$ complex roots that 
occur in conjugate pairs. As $p$ increases $F_p(x)$ is approximately 
constant over an increasing interval around $x=1/2$ and as 
$p\rightarrow\infty$, $F_{p}(x)=1$ for all $x$ except at $x=0,1$. 
(For example, $1-F_{p=16}(x)\geq 0.1$ only for $x\leq 0.035$ and 
$1-x\leq 0.035$.) Thus, $F_p(x)$ with large $p$ provides an excellent 
non-linear generalization of the linear-drift model without suffering from 
its limitations. We note that at finite $p$ 
Eq.(\ref{eq:nldriftv}) describes a memristive system~\cite{chua76,strukov} 
that is equivalent to an ideal memristor~\cite{chua71,strukov} when 
$p\rightarrow\infty$ or when the linear-drift approximation is applicable. 
It is instructive to compare the results for large $p$ with those for 
$p=1$, $F_{p=1}(x)=4x(1-x)$, when the window function imposes a non-linear 
drift over the {\it entire region} $0\leq w\leq D$.~\cite{strukov} For $p=1$ 
it is possible to integrate Eq.(\ref{eq:nldriftv}) analytically and we obtain 
\begin{equation}
\label{eq:nlwidth}
w_{p=1}(q)=w_0\frac{D\exp{(4\eta q(t)/Q_0)}}
{D+w_0\left[\exp{(4\eta q(t)/Q_0)}-1\right]}.
\end{equation} 
As expected, when the suppression at the boundaries is taken into account, 
the size of the doped region satisfies $0\le w(t)\le D$ for all $t$ and 
$w(t)$ asymptotically approaches $D (0)$ when $\eta=+1 (-1)$. For $p>1$, we 
numerically solve Eq.(\ref{eq:nldriftv}) with Kirchoff's voltage law applied 
to an ideal MCL circuit 
\begin{equation}
\label{eq:circuit}
L\frac{di}{dt}+M(q(t))i(t)+\frac{q(t)}{C}=v(t),
\end{equation}
using the following simple algorithm 
\begin{eqnarray}
\label{eq:wupdate}
w_{j+1}& = & w_j+\eta\frac{\mu_D\ron}{D}F\left(\frac{w_j}{D}\right)i_j,
\epsilon_t\\
\label{eq:iupdate}
i_{j+1}& = &i_j+\frac{\epsilon_t}{L}\left[v_j-\frac{q_j}{C}-
M(w_{j+1})i_j\right],\\
\label{eq:qupdate}
q_{j+1}& = & q_j+ i_{j+1}\epsilon_t.
\end{eqnarray}
Here, $\epsilon_t$ is the discrete time-step and $w_j,i_j$ and $q_j$ stand for 
the doped-region width, current, and charge at time $t_j=j\epsilon_t$ 
respectively. The algorithm is stable and accurate for small 
$\epsilon_t\leq 10^{-2}t_0$. 

\begin{figure}[h]
\begin{center}
\vspace{-12cm}
\includegraphics{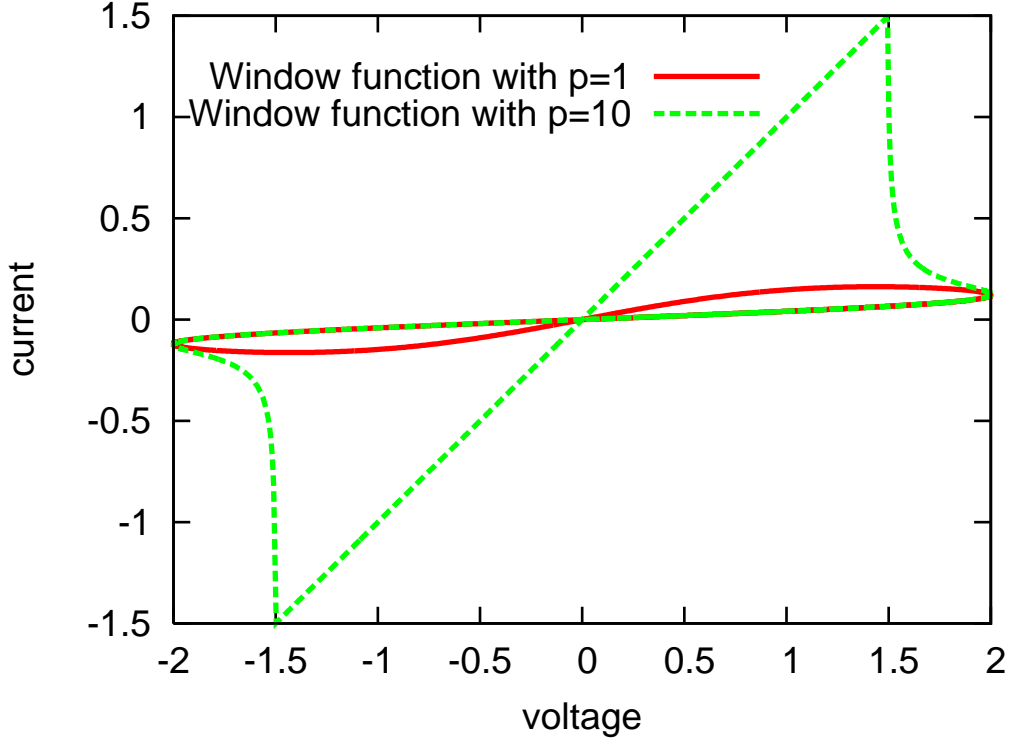}
\caption{\label{fig:nlmemristor}
Theoretical {\it i-v} curves for a memristor with (realistic) dopant drift 
modeled by window functions $F_p(x)=1-(2x-1)^{2p}$ with $p=1$ 
(red solid) and $p=10$ (green dashed), in the presence of an external voltage 
$v(t)=2v_0\sin(\omega_0 t/2)$. The memristor parameters are $w_0/D=0.5$ 
and $\roff/\ron=50$. We see that the memristive behavior is enhanced at 
$p=10$. The slope of the {\it i-v} curves at small times is the same, 
$\ro^{-1}$, in both cases whereas the slope on return sweep depends on the 
window function. For large $p$, the return-sweep slope is 
$\ron^{-1}=1\gg\ro^{-1}$ and it corresponds to a fully doped memristor.}
\end{center}
\end{figure}

\begin{figure}[h]
\begin{center}
\includegraphics{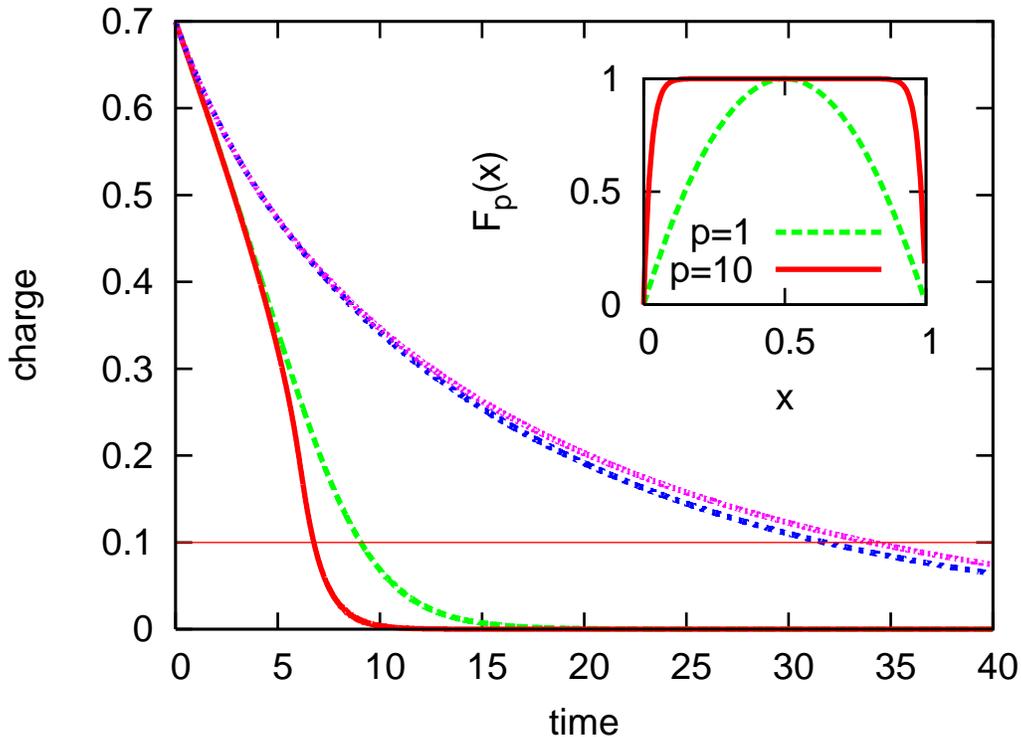}
\caption{\label{fig:nldischarge}
Theoretical {\it q-t} curves for an ideal MC circuit with non-linear 
dopant drift modeled by window functions $F_p(x)$ with $p=1$ and $p=10$ 
shown in the inset. The green dashed ($\eta=+1$) and the blue dash-dotted 
($\eta=-1$) correspond to $p=1$ window function. The red solid 
($\eta=+1$) and the magenta dotted ($\eta=-1$) correspond to the $p=10$ 
window function. The horizontal line at $q/Q_0=0.1$ is a guide to the eye. 
The memristor parameters are $w_0/D=0.5$ and $\roff/\ron=20$. The 
initial charge on the capacitor is $q_0/Q_0=0.7$ and $C/C_0=1$. We see that 
the memristive effect is enhanced for large $p$ when $\eta=+1$. Hence, 
for large $p$ the two decay time-scales associated with $\eta=+1$ 
(red solid) and $\eta=-1$ (magenta dotted) can differ by a factor of 
$\ro/\ron\gg 1$. Fitting the experimental data to these results can 
determine the nature of dopant drift in actual samples.}
\end{center}
\end{figure}

Figure~\ref{fig:nlmemristor} compares the theoretical {\it i-v} results for 
a single memristor with two models for the dopant drift: a $p=1$ model with 
non-uniform drift over the entire memristor (red solid) and a $p=10$ model 
in which the dopant drift is heavily suppressed only near the boundaries 
(green dashed). We see that as $p$ increases, beyond a critical voltage the 
memristance drops rapidly to $\ron$ as the entire memristor is doped. 
Figure~\ref{fig:nldischarge} shows theoretical results for a discharging 
ideal MC circuit obtained using two models: one with $p=1$ 
(green dashed for $\eta=+1$ and blue dash-dotted for $\eta=-1$) and the 
other with $p=10$ (red solid for $\eta=+1$ and magenta dotted for $\eta=-1$). 
The corresponding window functions $F_p(x)$ are shown in the inset. We 
observe that the memristive behavior is enhanced as $p$ increases, 
leading to a dramatic difference between the decay times of a single MC 
circuit when $\eta=+1$ (red solid) and $\eta=-1$ (magenta dotted). 
Fig.~\ref{fig:nldischarge} also shows that fitting the experimental data to 
these theoretical results can determine the window function that best 
captures the realistic dopant drift for a given sample. 

The properties of ideal MC and ML circuits with an arbitrary voltage are 
obtained by integrating Eqs.(\ref{eq:nldriftv}) and (\ref{eq:circuit}) 
using the algorithm described above. However, as the discussion in 
Sec.~\ref{sec:intro} shows, these circuits significantly differ from their 
ideal RC and RL counterparts only at low frequencies.   


\section{Oscillations and damping in an MCL Circuit}
\label{sec:mcl}

In this section, we discuss the last remaining elementary circuit, namely 
an ideal MCL circuit. First let us recall the results for an ideal RCL 
circuit.~\cite{ugbooks} For a circuit with no voltage source and an initial 
charge $q_0$, the time-dependent charge on the capacitor is given by 
\begin{eqnarray}
\label{eq:damped}
q(t) & =&
\left\{\begin{array}{l@{\qquad}l}
q_0 e^{-t/2\tau_{RL}}\cos(\tilde{\omega}t) & \tilde{\omega}^2>0 \\
q_0 e^{-t/2\tau_{RL}}\cosh(|\tilde{\omega}|t) & \tilde{\omega}^2<0\\ 
\end{array}\right.
\end{eqnarray}
where $\tilde\omega^2=\omega_{LC}^2-(2\tau_{RL})^{-2}>0$ defines an 
underdamped circuit and $\tilde{\omega}^2<0$ defines an overdamped circuit. 
The two results are continuous at $\tilde\omega=0$ (critically damped 
circuit). Thus, an RCL circuit is tuned through the critical damping when 
the resistance in the circuit is increased beyond $R_c=2\sqrt{L/C}$. 

\begin{figure}[h]
\begin{center}
\vspace{-12cm}
\scalebox{1.00}{\includegraphics{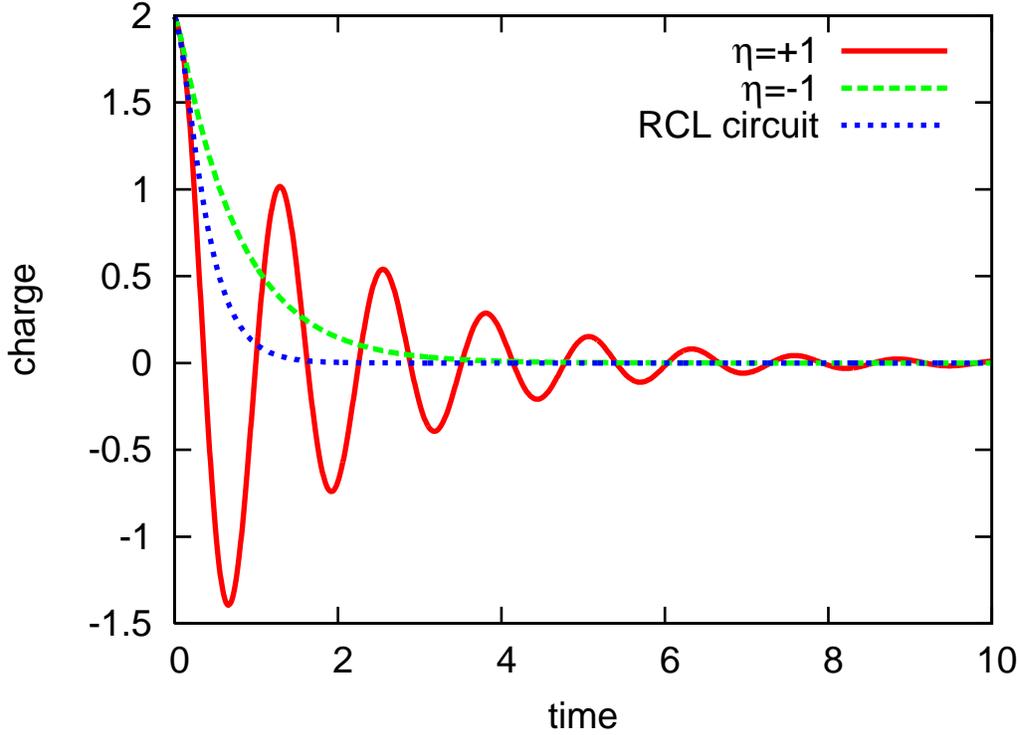}}
\caption{\label{fig:mcl}
Theoretical {\it q-t} curves for an ideal discharging MCL circuit 
modeled using the window function for $p=50$. The circuit parameters are 
$w_0/D=0.5$, 
$\roff/\ron=20$, $L/L_0=1$, $C/C_0=0.04$, and $q_0/Q_0=2$. The initial 
resistance $\ro=10.5$ implies that the corresponding ideal RCL circuit, with 
$\omega_{LC}=1/\sqrt{LC}\sim\ro/2L$, is close to critically damped. When 
$\eta=+1$ (red solid) we see that the MCL circuit is underdamped, whereas 
when $\eta=-1$ (green dashed) it is overdamped. Result for the RCL circuit 
with the same initial resistance $\ro$ is shown in blue dotted line. Thus, 
a single MCL circuit can be driven from overdamped to underdamped behavior by 
simply exchanging the $\pm$ plates on the capacitor.}
\end{center}
\end{figure}

The non-linear differential equation describing an MCL circuit, 
Eq.(\ref{eq:circuit}), cannot be solved analytically due to the 
$q$-dependent memristance. Figure~\ref{fig:mcl} shows theoretical {\it q-t} 
curves for a {\it single MCL circuit} obtained by numerically integrating 
Eqs.(\ref{eq:nldriftv}) and (\ref{eq:circuit}) using $p=50$ window function. 
When $\eta=+1$ (red solid) the circuit is underdamped because as the 
capacitor discharges the memristance reduces from its initial value $\ro$. 
When $\eta=-1$ (dashed green), the discharging capacitor {\it increases} 
the memristance. Therefore, when $\eta=-1$ the MCL circuit is overdamped. For 
comparison the blue dotted line shows the theoretical {\it q-t} result for 
an ideal RCL circuit with resistance $\ro$ that is chosen such that the 
circuit is close to critically damped, $\ro\sim 2\sqrt{L/C}$. 
Fig.~\ref{fig:mcl} implies that if we exchange the $\pm$ plates of the 
capacitor in an MCL circuit, the charge will decays rapidly or oscillate. 
This property is unique to an MCL circuit and arises essentially due to the 
memristive effects.

\begin{figure}[h]
\begin{center}
\vspace{-12cm}
\includegraphics{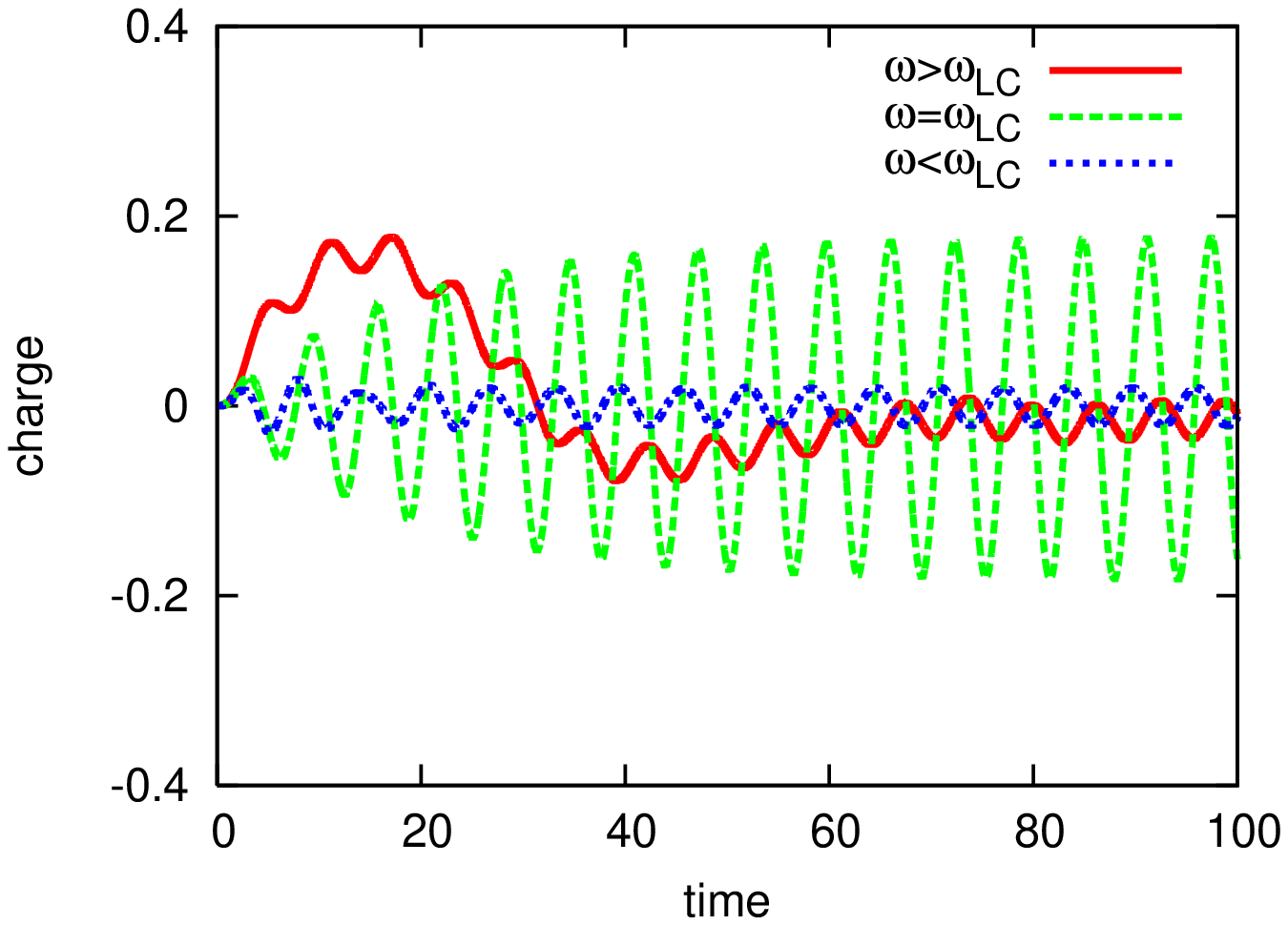}
\caption{\label{fig:mcldriven}
Theoretical {\it q-t} curves for an ideal MCL circuit driven by an AC 
voltage $v(t)=v_0\sin(\omega_0 t)$ with $\eta=+1$. The circuit 
parameters $w_0/D=0.5$, $\roff/\ron=10$, $L/L_0=50$ are fixed. The 
capacitance is $C/C_0=2$ (red solid), $C/C_0=0.02$ (green dashed), and 
$C/C_0=0.01$ (blue dotted). We see that for $\omega_{LC}<\omega_0$, the 
amplitude of the transient effects is comparable to the maximum amplitude 
that occurs at resonance, and that the memristive effect disappears in 
the steady-state solution.} 
\end{center}
\end{figure}

For completeness, we briefly discuss the behavior of an MCL circuit driven 
by an AC voltage source $v(t)=v_0\sin(\omega t)$, with zero initial charge 
on the capacitor. For an ideal RCL circuit, the steady-state charge $q(t)$ 
oscillates with the driving frequency $\omega$ and amplitude 
$v_0/L\sqrt{\left(\omega^2-\omega_{LC}^2\right)^2+(\omega/\tau_{RL})^2}$. 
For a given circuit, the maximum amplitude $v_0\sqrt{LC}/R$ occurs at 
resonance, $\omega=\omega_{LC}$ and diverges as 
$R\rightarrow 0$.~\cite{ugbooks} Fig.~\ref{fig:mcldriven} shows theoretical 
{\it q-t} curves for an ideal MCL circuit with $\eta=+1$ driven with 
$v(t)=v_0\sin(\omega_0t)$. The red solid line corresponds to low LC 
frequency $\omega_{LC}=0.1\omega_0$, the dashed green line corresponds to 
resonance, $\omega_0=\omega_{LC}$, and the dotted blue line corresponds to 
high LC frequency $\omega_{LC}=\sqrt{2}\omega_0$. We find that irrespective 
of the memristor polarity, the memristive effects are manifest only in the 
transient region. We leave it as an exercise for the Reader to explore the 
strong transient response for $\omega_{LC}<\omega$ and compare it with the 
steady state response at resonance $\omega_{LC}=\omega$.    


\section{Discussion}
\label{sec:disc}

In this tutorial, we have presented theoretical properties of the fourth 
ideal circuit element, the memristor, and of basic circuits that include a 
memristor. In keeping with the revered tradition in physics, the existence 
of an ideal memristor was predicted in 1971~\cite{chua71} based purely on 
symmetry argument~\cite{symmetry}; however, its experimental 
discovery~\cite{oldworkprogram,thakoor,ero1,ero2,strukov} 
and the accompanying elegant physical picture~\cite{strukov,pershin} took 
another 37 years. The circuits we discussed complement the standard RC, RL, 
LC, RCL circuits, thus covering all possible circuits that can be formed 
using the four ideal elements (a memristor, a resistor, a 
capacitor, and an inductor) and a voltage source. We have shown in this 
tutorial that many phenomena - the change in the discharge rate of a 
capacitor when the $\pm$ plates are switched or change in the current in a 
circuit when the battery terminals are swapped - are attributable to a 
memristive component in the circuit.~\cite{miao,pershin} In such cases, a 
real-world circuit can only be mapped on to one of the ideal circuits with 
memristors. 

The primary property of the memristor is the {\it memory of the charge that 
has passed through it}, reflected in its effective resistance $M(q)$. 
Although the microscopic mechanisms for this memory can be 
different,~\cite{strukov,pershin} dimensional analysis implies that the 
memristor size $D$ and mobility $\mu_D$ provide a unit of magnetic flux 
$D^2/\mu_D$ that characterizes the memristor.  
Although the underlying idea behind a memristor is straightforward, its 
nano-scale size remains the main challenge in creating and experimentally 
investigating basic electrical circuits discussed in this article. 

We conclude this tutorial by mentioning an alternate possibility. It is 
well-known that an RCL circuit is 
equivalent~\cite{ugbooks} to a 1-dimensional mass+spring system in which the 
position $y(t)$ of the point mass is equivalent to the charge $q(t)$, the 
mass is $L$, the spring constant is $1/C$, and the viscous drag force is 
given by $F(v)=-\gamma v$ where $\gamma=R$. Therefore, a memristor is 
equivalent to a viscous force with a $y$-dependent drag coefficient, 
$F_M=-\gamma(y)v$. Choosing $\gamma(y)=\gamma_0-\Delta\gamma y/A$, where 
$A$ is the typical stretch of the spring, will create the equivalent of an 
MCL circuit.Since a viscous force naturally occurs in fluids, a vertical 
mass+spring system in which the mass moves inside a 
fluid with a large vertical viscosity gradient can provide a macroscopic 
realization of the MCL circuit.


\begin{acknowledgments}
It is a pleasure to thank R. Decca, A. Gavrin, G. Novak, and K. Vasavada for 
helpful discussions and comments. 
This work was supported by the IUPUI Undergraduate 
Research Opportunity Program (UROP). S.J.W. acknowledges a UROP Summer 
Fellowship. 
\end{acknowledgments}



\newpage
\section*{Figures}

\begin{thebibliography}{99}
\bibitem{caveat1} The linear-drift model is valid only when 
$0\leq w(t)\leq D$ for all $t$. This constraint provides limits on the 
flux $\phi$, the initial capacitor charge $q_0$ or the DC voltage $v_0$, and 
the initial current $i_0$. We compare linear and non-linear dopant drift 
models in Sec.~\ref{sec:nldrift}.
\bibitem{symmetry} The displacement-current in Maxwell's equations, a 
positron, and a magnetic monopole are a few historical examples. The 
first two have been experimentally observed, while the third one remains 
elusive.
\end{thebibliography}
\end{document}